\documentclass{aa}
\usepackage{graphicx}
\usepackage{txfonts}
\begin{document}
   \title{Spatial simulations of the Kelvin-Helmholtz instability in astrophysical jets}
   \subtitle{A partial stabilization mechanism for weakly magnetized transonic flows}
      
   \author{M. Viallet \and H. Baty}
   
   \offprints{M. Viallet, viallet@astro.u-strasbg.fr}
   
   \institute{Observatoire Astronomique - ULP \& CNRS - 11, rue de l'Universit\'e - 67000 Strasbourg - France\\
}
              
   \date{Received ; accepted}
   

\abstract
{}
{The long term magnetohydrodynamic stability of magnetized transonic/supersonic jets is numerically investigated using a spatial approach. We focus on two-dimensional linearly-unstable slab configurations where the jet is embedded in a flow-aligned uniform magnetic field of weak amplitude. We compare our results with previous studies using a temporal approach where longitudinally periodic domains were adopted.}
{The finite-volume based versatile advection code is used to solve the full set of ideal compressible MHD equations. We follow the development of Kelvin-Helmholtz modes that are driven by a white noise perturbation continuously introduced at the jet inlet.}
{No noticeable difference is observed in spatial simulations versus analogous temporal ones during the linear and early non-linear evolution of the configuration. However, in the case of transonic flows, a different long-term scenario occurs in our spatial runs. Indeed, after the large-scale disruption of the flow, a sheath region of enhanced magnetic field encompassing the jet core forms along the whole flow. This provides a partial stabilization mechanism leading to enhanced stability for later times, which is almost independent of the initial magnitude of the magnetic field. The implication of this mechanism for the stability of astrophysical jets is discussed.}
{}

\keywords{instabilities - magnetohydrodynamics (MHD) - ISM : jets and outflows - methods : numerical - plasmas}

\maketitle
\section{Motivation}

Astrophysical jets are observed to propagate over very long distances with respect to their radial 
extent, typically up to one thousand times the jet radius.
This is, for example, the case of many jets emanating from young stellar objects (YSO) and  active 
galactic nuclei (AGN). These observations seem to be in disagreement with the magnetohydrodynamic (MHD) stability theory and numerical simulations, which predict the strong 
development of instabilities that consequently threaten the jet's collimation. 
Particularly important is the Kelvin-Helmholtz (KH) instability, typical of shear flows, which is shown 
to disrupt the jet on a few sound transit timescales (see reviews by Birkinshaw  \cite{birkinshaw}, and Ferrari 1998). This is mostly evident in purely hydrodynamic simulations, where supersonic flows undergo a strong turbulent disruption driven by the KH modes (Bodo et al. 1995, 1998).

Among the possible mechanisms that are likely to have a stabilizing effect is the role of magnetic 
fields. Indeed, the presence of a large-scale magnetic field in accretion-disk-jet systems
is now well established from observations (Ray et al. 1997; Pushkarev et al. 2005; Gabuzda et 
al. 2004) and jet-launching models (Casse \& Keppens 2002, 2004 and references therein).
The magnetic field is probably too weak to prevent or strongly weaken the linear development of the 
KH instability (Baty 2005; Appl \& Camenzind 1992). However, the evolution of the
MHD KH instability  allows a much richer complexity in the non-linear phase, compared to its purely 
hydrodynamic counterpart. For example, a weakly magnetized shear flow configuration has
three dynamically different subregimes depending on the value of the Alfv\'en Mach number (Baty et 
al. 2003; Jones et al. 1997). A recent numerical investigation showed that the overall
scenario, even if it is enriched by magnetic reconnection events and large-scale coalescence 
effects, leads to quick disruption in MHD as well (Baty \& Keppens 2006, referred to as Paper I in this study). The latter result is valid for the non-linear evolution of uniformly magnetized jets having widely different plasma-beta, sonic/Alfv\'enic Mach numbers, and shape of the flow profile.
However, a temporal approach with periodic boundary conditions imposed at the inflow/outflow 
boundaries is used in Paper I. Indeed, this temporal procedure is preferable when one wants
to focus on long-term temporal evolution, as it allows one to compute the time evolution of a jet 
section at relatively low numerical cost. On the other hand, the temporal approach has
several limitations (Baty et al. \cite{baty2003}). The most important one is that the interaction 
between different parts of the jet is not well taken into account, as the convective effect
of the mean flow is ignored in periodic configurations (even in cases of extended domains). 

Our aim is to investigate the development  of the KH instability using a spatial approach, where we follow the evolution of a perturbation introduced at the jet inlet of a non-periodic configuration. This implies the use of an extended numerical domain, as the study is limited by the transit time through the jet configuration. We adopt similar physical parameters for the jet as those assumed in Paper I, in order to make a close comparison with results obtained using a temporal approach. We consider only two dimensional (2D) configurations, mainly because the numerical cost of a similar study in three dimensions (3D) is beyond our current computational capabilities.

We mainly focus here on weakly magnetized transonic flows. This is a relevant regime for studying the early phase of jet propagation. Indeed, before reaching the asymptotic region (where the flow is believed to be supersonic), the jet is accelerated in an intermediate region where it must survive instabilities. And the latter intermediate regime is of prime importance for KH modes, as the spatial growth rates are larger for transonic flows vs supersonic flows. A selected supersonic jet case is nevertheless investigated to address the generality/limitation of our transonic results. Previous numerical studies have already adopted a spatial approach (e.g. Hardee et al. \cite{hardee1}; Hardee et al. 1995; Micono et al. \cite{micono}). However, the latter studies focus mainly on relatively strongly magnetized jets and describe the whole jet in order to make comparisons with observational characteristics; and for this purpose they necessarily sacrifice resolution in the description of modes (especially in 3D). Finally, note that we do not deal here with magnetic current-driven instabilities, which are driven by the presence on an electrical current density parallel to the magnetic field in 3D helical field geometry.

This paper is organized as follows. We present the jet model and numerical setup used in this study in Sect. 2. In Sect. 3, we show the numerical results, which are split into an overview of a typical transonic flow run, a comparison with temporal simulations, an in-depth study of a partial stabilization mechanism observed in our simulations, and an overview of a supersonic run. Finally, conclusions are drawn in Sect. 4.

\section{Jet model and numerical setup}
\label{setup}

In the present study, we use the full set of ideal compressible MHD equations, as we consider 
perfectly conducting plasmas having very large kinetic and magnetic Reynolds numbers.
This description is a suitable proxy for astrophysical jets environments. Thus, we rely on the 
inherent numerical resistivity and viscosity (due to scheme discretizations) to mimic the dissipative
processes. It is generally admitted that this approach provides an adequate model for subgrid-scale 
dissipation (see Paper I, and the discussion in Jones et al. 1997). However, a convergence
study by repeating the simulations using different grid resolutions is recommended.

\begin{table}
\begin{minipage}[t]{\columnwidth}
\caption{Numerical and physical parameters of our simulations. Run 2b corresponds to the 
numerical linear convergence study and is done on a smaller domain.}
\label{tab_par}
\centering
\renewcommand{\footnoterule}{}  
\begin{tabular}{cccccc}
\hline \hline
Run & $M_s$   & $M_A$ & $M_f$ & Domain & Resolution  \\
\hline
   1  & 1 &   5  &   0.98  &  [0,40]x[-8,8] &1200x480\\
   2  & 1 &   7  &   0.99  &  [0,40]x[-8,8] &1200x480\\
   2b& 1 &   7  &   0.99  &  [0,20]x[-2,2] & 200x40, 300x60, \\
       &    &       &            &  &600x120, 800x160\\
       &    &       &            &   &1200x240,1600x320\\
   3  & 1 &  10 &   0.99  &  [0,40]x[-8,8] & 1200x480\\
   4  & 1 &  14 &   0.997  &  [0,40]x[-8,8] &1200x480\\
   5  & 1 &    5 &   0.98  &  [0,80]x[-16,16] & 1200x480\\
   6  & 1 &    7 &   0.99  &   [0,80]x[-16,16] & 1000x600\\
   7  & 1 &  14 &   0.997 &  [0,80]x[-20,20] & 1000x600\\
   8  & 3 &   7  &   2.75  & [0,80]x[-16,16] & 1000x600\\
\hline
\end{tabular}
\end{minipage}
The numerical domain is rectangular, of size $[0, L_x]\times[-L_y/2,L_y/2]$. In transonic runs, the amplitude of the jet velocity is $V_0 = 1.29$ and the initial magnetic field intensity is varied. In the supersonic run, $V_0 = 3.87$.
\end{table}

The set of MHD equations is solved in 2D Cartesian geometry, with  the direction of flow 
propagation being along the $x$ axis. The transverse coordinate is $y$. We consider a jet having
an initial background flow profile:

\begin{equation}
  \label{eq:profile}
  V_x(y) = \frac{V_0}{2}\Big ( 1- \tanh F \big (\frac{|y|}{R_j}-\frac{R_j}{|y|} \big ) \Big )  ,
\end{equation}
\noindent

\noindent
where $V_0$ is the velocity amplitude of the jet. The parameters $R_j$ and $F$ control the shape 
of the velocity profile; $R_j$ is the jet radius, and $F$ determines the thickness $\epsilon$ of the shear flow ($\epsilon = R_j/F$, see Paper I). For convenience, we also introduce the half-thickness of the vorticity layer $a=\epsilon/2$. In this study, we use a flow profile similar to the $Pr2$ case studied in Paper I ; more explicitly we take $R_j = 0.5$, and $F = 2$ (thus $a=0.125$ in our case, which differs from Paper I where $a=0.05$ for all profiles). We consider a medium with initially uniform temperature $T_0=1$ and density $\rho_0=1$, thus defining our normalization; the sonic speed $c_s = (\gamma P_0/\rho_0)^{1/2}$ is then equal to $1.29$ in our units (with the adiabatic index $\gamma=5/3$). Consequently, the time needed for sound waves to travel through the jet radius is $R_j/c_s=0.3876$ in our units.

The flow is embedded in an initially uniform longitudinal magnetic field of amplitude $B_0$. The 
parameters used in the different runs are listed in Table \ref{tab_par}, where we take  the following
definitions for the sonic and Alfv\'enic Mach numbers, $M_s = V_0/c_s$ and $M_A = V_0/V_A$ 
($V_A = B_0/\sqrt{\rho_0}$ is the Alfv\'en velocity). The fast Mach number is defined as $M_f = V_0/\sqrt{c_s^2+V_A^2}$.As already emphasized, we focus on transonic cases ($M_s=1$). However, a supersonic case $M_s=3$ is also considered for comparison and for the sake of completeness in Sect. \ref{supersonic}. Relatively weak magnetic field amplitudes are taken with $M_A$ in the range $[5-14]$ which correspond to values of the so-called ``weak field regime" or ``disruptive regime" (e.g. Jones et al. \cite{jones}; Baty et al. 2003). Indeed, higher $M_A$ values seem to be unrealistic for astrophysical jets and lower values lead to stable configurations (also probably unrealistic).

   \begin{figure*}
   \centering
      \includegraphics[width=17cm]{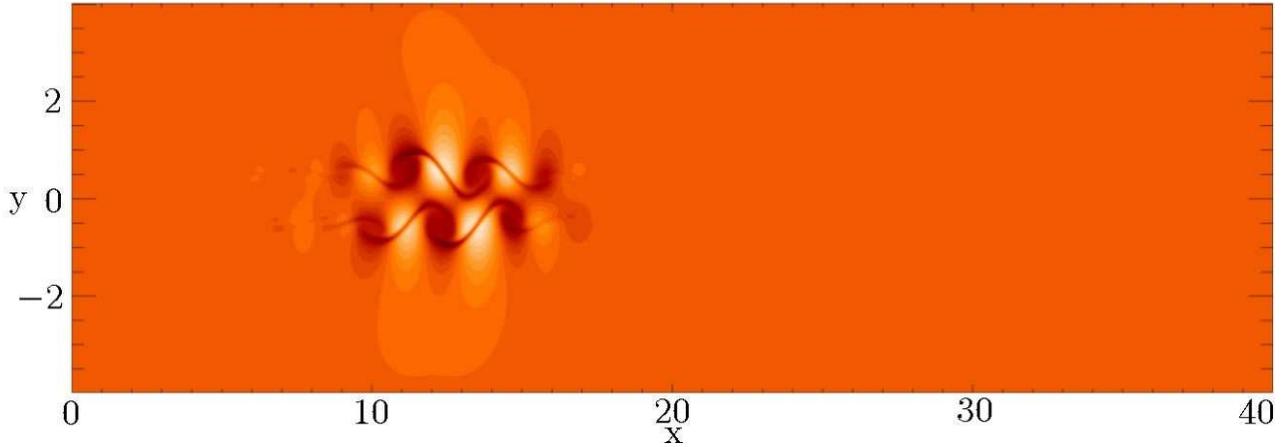}
      \caption{Density distribution in the x-y plane obtained at time $t = 16$ for run 2 (see 
Table \ref{tab_par}). Dark regions correspond to low density values. A linear scale is used with density values ranging 
from 0.6 to 1.2. The unit of length is the jet diameter.}
       \label{t1}
   \end{figure*}

   \begin{figure*}
   \centering
      \includegraphics[width=17cm]{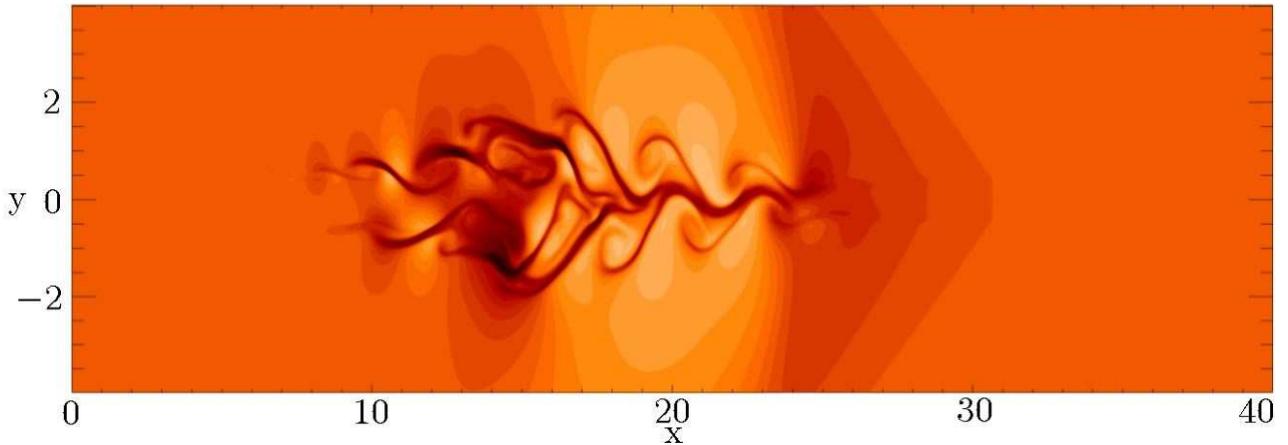}      
      \caption{Same as Fig. \ref{t1}, at $t  = 22$.}
       \label{t2}
   \end{figure*}

   \begin{figure*}
      \centering
      \includegraphics[width=17cm]{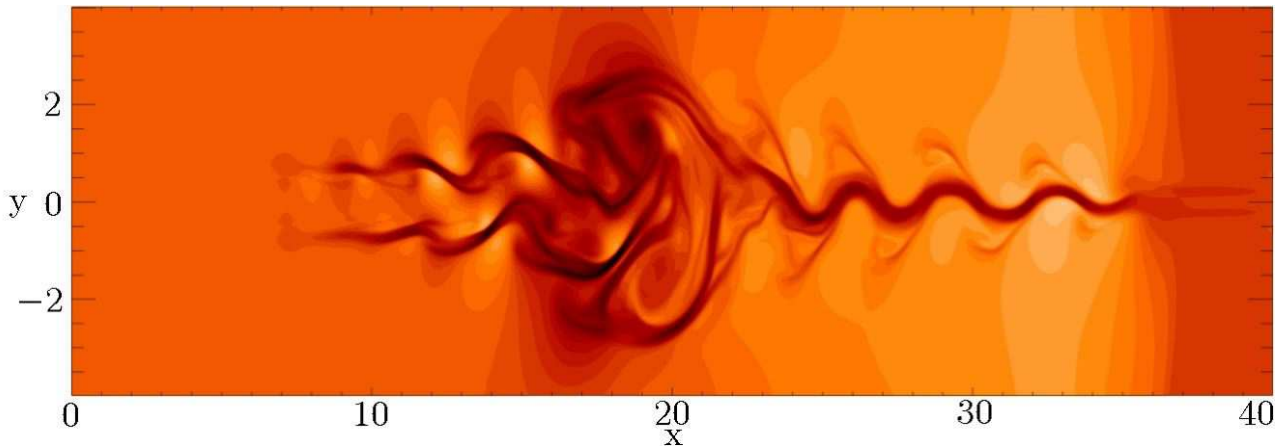}
      \caption{Same as Fig. \ref{t1}, at $t  = 34$.}
      \label{t3}
   \end{figure*}

We use the general finite-volume based versatile advection code VAC (T\'oth \cite{toth}) for our 
simulations. The MHD equations are solved by selecting the explicit one-step total variation
diminishing (TVD) scheme with minmod limiting (Colella \& Woodward 1984; Harten 1983). This is a 
second-order accurate shock-capturing method using a Roe-type approximate Riemann
solver. Our VAC simulations also apply a projection scheme at every time-step to remove any numerically generated divergence of the magnetic field up to a predefined accuracy. Spatial simulations require an extended numerical domain; Table  \ref{tab_par} summarizes the different 
domain sizes used. In short, we employed two different length values for the numerical 
domain: a ``short" one with $L_x = 40$ having a relatively high spatial resolution, and a ``large" one with $L_x=80$ having a lower resolution to study the large-scale stability of the jet.
As an initial condition, the jet profile (1) was set up across the whole computational domain.
The jet continuously enters the domain by its left boundary and free outflow conditions were used 
on all other boundaries (by imposing a zero gradient for every variable). Lateral boundaries were 
taken far enough away from the jet flow to avoid spurious effects. Thus, at the jet inlet (left 
boundary at $x=0$), the background velocity profile (\ref{eq:profile}) is maintained in time. 
Moreover, to excite the unstable KH modes, we continuously perturbed the jet inlet (at $x=0$) by adding a small amplitude transverse velocity, which is taken to be one percent of the magnitude of the background flow. Specifically, as Paper I, we imposed a white noise perturbation by using a random number generator. This is a valid procedure for selecting the linearly fastest growing mode as shown by Zhao et al. (1992). Concerning the other physical variables at this boundary, we imposed fixed conditions. Indeed, the real physical conditions are not known and this latter, somewhat arbitrary, choice is convenient for our numerical procedure.


\section{Results}

Before examining in detail the simulation results, it is informative to have an overview of the typical 
evolution of a transonic case (run 2 in Table 1).

\subsection{Overview of a transonic run}
\label{overview}

Clearly, two simulations will always differ in their details, in particular due to the random 
character introduced by the perturbation. However, the following scenario can be inferred from our 
simulations.
Figures \ref{t1}, \ref{t2} and \ref{t3} show the first stages of the transonic run 2 (see Table 1). First, one can clearly see in the density snapshots of Fig. 1 the initial development of a few unstable KH wavelengths, which take the form of two rows of vortices located on the two shear layers of the mean flow. We term this set of growing instabilities that are convected along the flow direction, the \emph{instability train}. Later, these vortical structures undergo pairing-merging events (due to an inverse cascade effect towards large scales, as described in Paper I) accompanied by magnetic reconnection, and finally a strong disruption (Figs. \ref{t2}, \ref{t3}). This scenario is
characteristic of the disruptive regime, and is  expected from temporal studies. In the next section, a  detailed comparison is made with results obtained in a similar temporal simulation, to quantify the convective effect in our spatial run.

\begin{figure}
   \centering
   \includegraphics[angle=-90,width=9cm]{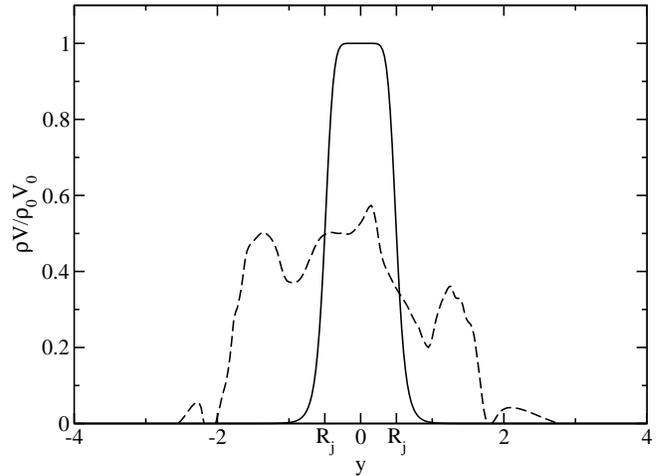}
   \caption{One dimensional cut of the normalized x-momentum (momentum $\rho V$ divided by the initial momentum $\rho_0 V_0$)  profile, obtained at $x = 20$ (dashed line) and $x=0$ (plain line) in run 2 at $t=34$ (Fig. \ref{t3}).}
   \label{profils}
\end{figure}

The disruption occurring at t = 34 in Fig. \ref{t3} completely destroys the initial laminar structure of 
the jet's flow. This is illustrated in Fig. \ref{profils}, where the normalized x-momentum 
profile at the location of the large-scale disruption in Fig. \ref{t3} ($x \approx 20$ at $t=34$) and at the inlet is plotted.
As a consequence, we can estimate a maximum propagation distance over which the jet flow is able 
to maintain its coherence, that is roughly equal to 40 jet radii in our simulation.
This result is consistent with a value deduced from previous temporal studies, where a similar value 
for a disruption length was obtained by translating the typical disruption time
(Paper I). This is also in close agreement with the value deduced from the additional temporal 
run performed in the present study (see Sect. \ref{linear}).

In Figs. \ref{t2} and \ref{t3}, one can see two additional striking features. First, at the head of the 
instability train, one can see the formation of a propagation front (located at $x \approx 24$ in Fig. \ref{t2}, and at $x\approx37$ in Fig. \ref{t3}), which is found to travel in the downstream direction at approximately the jet's speed.
Behind this front, a sinusoidal density pattern is also formed (Fig. \ref{t3}, between $x\approx23$ and $x\approx37$). The latter structure arises from secondary instabilities which are driven
at the head of the instability train. This ``convective generation" of instabilities is discussed in 
Yamamoto et al. 1988. Indeed, the primary modes of the instability train can
non-linearly trigger secondary KH modes when convected by the mean flow, since the downstream 
medium is taken to be linearly unstable.
With our jet-like velocity profile, this leads to the formation of a single vortex-like street  (sinusoidal 
pattern) located in the jet core, as one can see in the density snapshot of Fig 3.
We verified that the latter phenomenon has no effect on the large-scale disruption of the jet 
(occurring far behind the front), and thus no consequence on the flow survival.
The second striking feature is the formation on the two edge layers of a trail structure, which is found to propagate in the upstream direction toward the jet's inlet (Fig. \ref{t3} between $x\approx7$ and $x\approx11$).
This trail is also visible at early times, between $x\approx8$ and $x\approx10$ in Fig. \ref{t2} (but 
less clearly). We showed that this particular phenomenon is associated with a local
magnetic amplification process (during the formation of the KH vortices of the instability train) which 
is able to reach the jet inlet at later times. A partial stabilization mechanism thus ensues as 
demonstrated in detail in Sect. \ref{retroaction_section}.

\subsection{Comparison with the temporal approach}
\label{linear}

We compare here the results obtained in our typical transonic spatial case (run 2) with 
results obtained in a similar temporal simulation.
Note that the physical parameters of the jet configuration are the same, and also that we use the 
same code, VAC.
The only difference comes from the numerical procedure, where periodic boundary conditions are 
assumed in the longitudinal direction for the temporal run (see Paper I for more details).
The periodic length of the domain is taken to be $L_x=8$, to allow a few wavelengths of the 
linearly fastest growing mode to fit the extent of the spatial domain
(see Fig. \ref{vortex_temp}).

Before examining the results in details, one needs to have in mind the main mathematical 
differences between the two approaches (spatial and temporal) for the description of linearly
growing instabilities in such flowing plasmas. The standard linear analysis assumes an infinitely long 
jet in equilibrium. 
The linear perturbations are thus conveniently expanded using the classical normal mode form:

\begin{equation}
f(x,y,t) = A(y)\exp \big [ i (kx - \omega t)\big ] ,
\label{eq:mode}
\end{equation}
for any perturbed physical quantity $f$.
Here, $\omega$ and $k$ are respectively the frequency and wavenumber of a given normal mode.
The linearized MHD equations then yield a dispersion relation:
\begin{equation}
D(k,\omega) = 0 ,
\label{eq:dispersion}
\end{equation}

\noindent which is specific to the jet configuration. The stability of normal modes is obtained by 
solving this dispersion equation.
This can be done in two ways.  First, in a temporal approach, a complex frequency $\omega = 
\omega_r + i\gamma$ and real wavenumber $k$ are assumed.
Consequently, $\gamma$ is the temporal linear growth rate of the corresponding wavenumber $k$. 
On the other hand, a spatial approach considers that the frequency $\omega$ is real and that the 
wavenumber $k =k_r + i\Gamma$ is now complex. As a consequence, the linear spatial growth rate 
is directly obtained by $\Gamma$, and $\Gamma^{-1}$ gives the corresponding characteristic 
growth length. This latter formalism better accounts for the convective nature of the KH instabilities,
which are not growing at a fixed point in space, but instead are convected along the fluid forming a 
spatially growing pattern.  Both approaches can be connected using the group velocity $v_g$
of the modes, with $\Gamma= \gamma/v_g$ (Drazin \& Reid 1981; Appl \& Camenzind 1992).
Moreover, as the modes under consideration in this study are weakly dispersive, one can also use 
the relation $\Gamma \simeq \gamma/v_p$ where $v_p$ is the phase velocity (Payne \& 
Cohn \cite{payne}; Appl 1996). 

  \begin{figure}
     \centering
     \resizebox{\hsize}{!}{\includegraphics{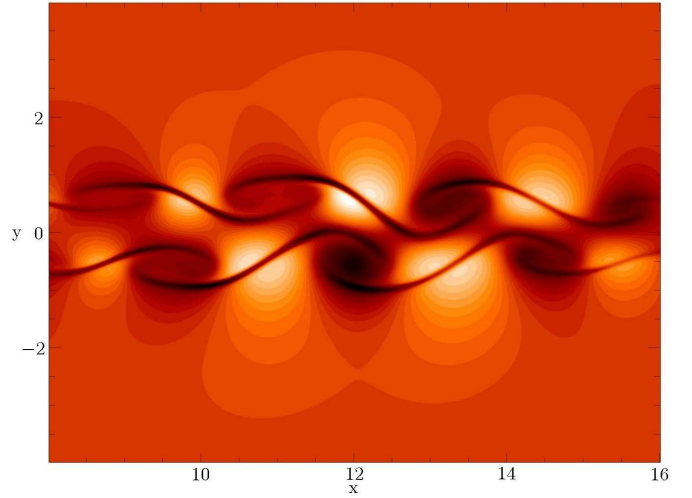}}
      \caption{Zoom on the instabilities train, taken from the spatial run 2 at $t = 16$ (Fig. \ref{t1}). A linear scale is used with density values ranging from 0.6 to 1.2. The unit of length is the jet diameter.}
      \label{vortex_spa}
   \end{figure}

\begin{figure}
   \centering
   \includegraphics[width=8.7cm]{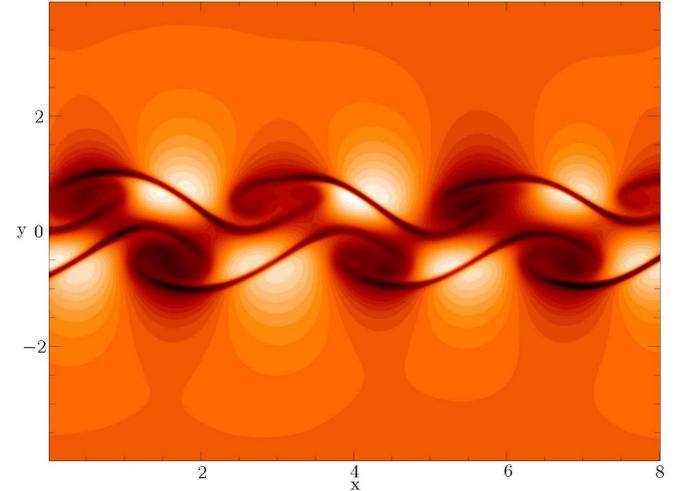}
      \caption{Density distribution for a temporal simulation obtained at a similar time as Fig. \ref
{vortex_spa} (see text). The unit of length is the jet diameter.}
         \label{vortex_temp}
   \end{figure}

Figure \ref{vortex_spa} shows a zoom of the instability train (plasma density) obtained in the typical 
transonic spatial simulation displayed in Fig. \ref{t1}. One can see that the latter snapshot
is very similar to the density map obtained from a temporal simulation (Fig. \ref{vortex_temp}). 
In both runs, a double layer of three vortices localized on the shear flow is clearly visible. The
longitudinal extent of each vortex is  $\lambda \approx 2$ in our units, which corresponds to a normalized wavenumber $ka \approx 0.4$. This is in excellent agreement with results obtained from
a linear stability analysis of the slab jet where the wavenumber of the linearly fastest growing mode is $ka \approx  0.4$, and does not depend significantly on the form factor $F$ of the jet (e.g. the growth rates curve in Fig. 1 of Paper I, where the maximum rate occurs for $k \approx 8$ with a profile having $a=0.05$). The wavelength of the fastest growing mode of the slab jet is also very close to the wavenumber obtained for a single shear layer embedded in a uniform magnetic field, where $ka \approx  0.41$ (Keppens et al. 1999; Miura \& Pritchett 1982). This is not surprising as the aspect ratio of the jet flow profile considered in this study is quite large ($R_j/a = 4 \gg 1$), leading thus to a rather weak interaction between both layers during the linear regime. A similar conclusion was previously reported by Min (1997).

In order to have a quantitative comparison of both approaches, we measured the growth rates by following the growth of the mean transverse kinetic energy: 
 
 \begin{equation}
  E_{ky}(x,t)=\frac{1}{L_y}\int_{-L_y/2}^{L_y/2}dy\frac{1}{2}\rho(x,y,t)v_y^2(x,y,t) ,
   \label{eq:eky}
\end{equation}

\noindent
which increases exponentially in time during the linear development of the instability. 

\begin{figure}
   \centering
   \includegraphics[width=9cm]{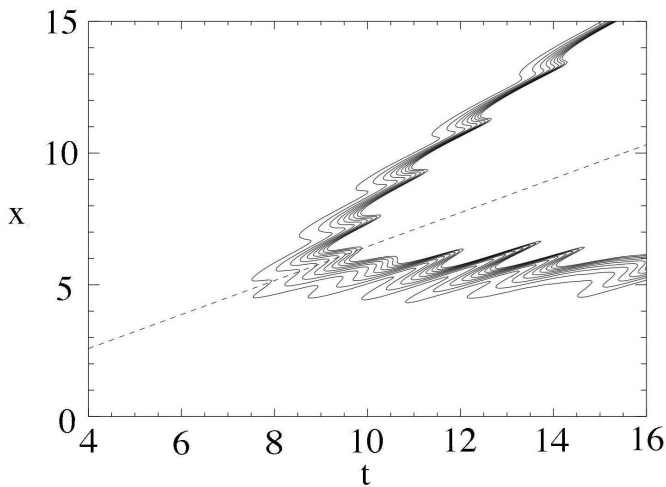}
      \caption{Isocontours of the mean transverse kinetic energy (see Eq. (\ref{eq:eky})) in the 
(t,x) plane. Ten linearly spaced isocontour values (between a minimum and a maximum) are 
considered. The dashed line is the characteristic curve corresponding to an instability starting to 
grow at $t=0$ and $x=0$.}
         \label{diagEky}
 \end{figure}
 
For the spatial run, one cannot easily separate the spatial and time dependencies. For example, 
Fig. \ref{diagEky} shows an isocontour diagram of the mean transverse kinetic energy
(\ref{eq:eky}) in the $(t,x)$ plane. The range of the isocontour values is taken well above the 
amplitude of the initial perturbation and sufficiently below the amplitude of saturation of the 
instabilities, in order to address the linear regime. The ten different isocontour values that are 
plotted in the diagram are taken to be equally spaced between the minimum and the maximum 
value.
Thus, the latter figure allows us to ``catch" the  growing instabilities as their mean transverse kinetic 
energy reach the level corresponding to the isocontour values.
Each of the 11 bump-like structures shown in Fig. \ref{diagEky} indicates the presence of a 
growing mode. Furthermore, the isocontours appear to become more and more
clustered (in the $(t,x)$ plane) as time goes on (recall that the isocontour values are linearly 
spaced). This is due to the exponential growth of the instabilities. In the same way, each ``bump" in 
successive isocontours is shifted according to the phase velocity of the different modes. 
One can deduce the phase velocity of the different modes simply by measuring the slope of the 
characteristic curve joining the different isocontour curves.
We estimated the phase velocity of the instabilities to be $V \approx 0.65$, which is close 
to the expected theoretical value of $V_p = V_0/2$ (e.g. Drazin \& Reid 1981).  
Moreover, a simple fitting of the transverse kinetic energy along the corresponding characteristic by 
an exponential function gives an estimate of the temporal growth rate.
It is important to note that instabilities situated below the characteristic curve $x=V_p t$ (corresponding to 
a mode growing from $x=0$ at $t=0$ and shown as a dashed line in Fig. \ref{diagEky}) are successively growing modes from the initial perturbation (situated at $x=0$). On the other hand, the instabilities situated above this characteristic curve are the secondary instabilities which are driven by the primary modes of the instability train (as discussed in the previous section).
As these secondary instabilities correspond to a non-linear driving phenomenon, we do not expect 
them to be correctly described by a linear theory. We deduce from Fig. \ref{diagEky} a temporal linear growth rate  $\gamma \approx 0.125V_0/(2a)$ (see also Fig. 8). A very similar value can be deduced from our additional temporal run and also from Paper I.

\begin{figure}
   \centering
   \includegraphics[angle=-90,width=9cm]{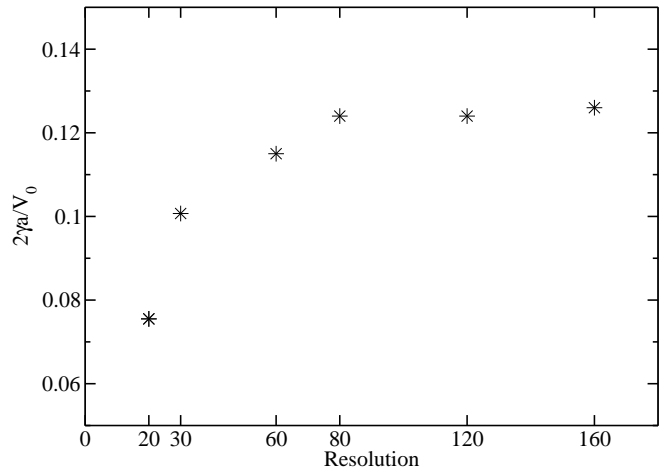}
      \caption{Normalized temporal growth rates obtained from spatial runs 2b (see Table 1) as a function of the number of grid points (in the longitudinal direction) used to describe one wavelength $\lambda \approx 2$ of the linearly fastest growing mode.}
         \label{res_gamma}
\end{figure}

Furthermore, we checked the numerical convergence of our spatial results during the linear 
regime by repeating run 2 at different resolutions (see run 2b in Table 1). As we are only interested 
in the linear regime, we can safely use a smaller numerical domain to reduce the numerical cost.
The measured temporal growth rates are plotted in Fig. \ref{res_gamma}, showing that using $60$ grid points per longitudinal wavelength is sufficient to achieve a good convergence.
Moreover, note that taking only $30$ grid points is still acceptable, while allowing a considerable 
reduction of the numerical cost. Such ``low resolution" run is used to investigate the long-term 
evolution (Sect. \ref{retroaction_section}).

In conclusion, our study does not show any noticeable difference in the linear development of the 
KH instability in the spatial vs temporal approaches. We checked that this is also true during the early non-linear stages of our simulations, where a similar scenario with pairing/merging events between adjacent vortices leading subsequently to a large-scale disruption occurs before $x\approx 40 R_j$. However, this is no longer true for the further non-linear evolution of the system, as we will see in the following section.

\subsection{Stabilizing mechanism}
\label{retroaction_section}
	
\subsubsection{Description}
\begin{figure*}
   \centering
   \includegraphics[width=18cm]{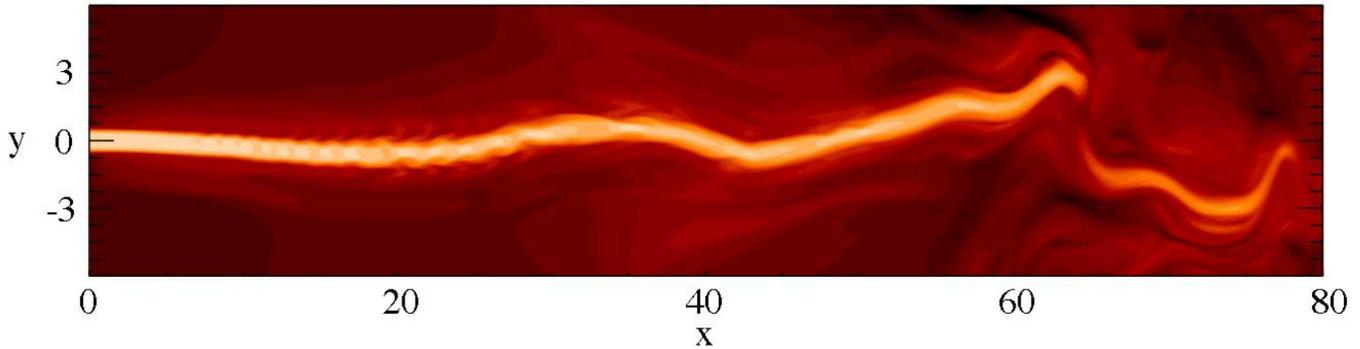}
      \caption{Axial velocity distributions from run 6 (same parameters as in run 2, but with a 
numerical domain twice as large) a long time after the large-scale disruption ($t = 320$). Dark 
regions correspond to low velocity values. A linear scale is used with values ranging from 0. 
to 1.29. The unit of length is the jet diameter.}
         \label{jetl80}
 \end{figure*}
	
At later times in a temporal simulation, the jet configuration is  able to reach a relaxed state which is 
a quasi-steady residual flow (Paper I). The final flow is thus drastically enlarged, invading the 
whole numerical domain, and most of the initial kinetic energy is transferred to the external medium. 
Consequently, the initial jet flow is not able to survive in temporal simulations.

 \begin{figure*}
    \centering
          \includegraphics[width=16cm]{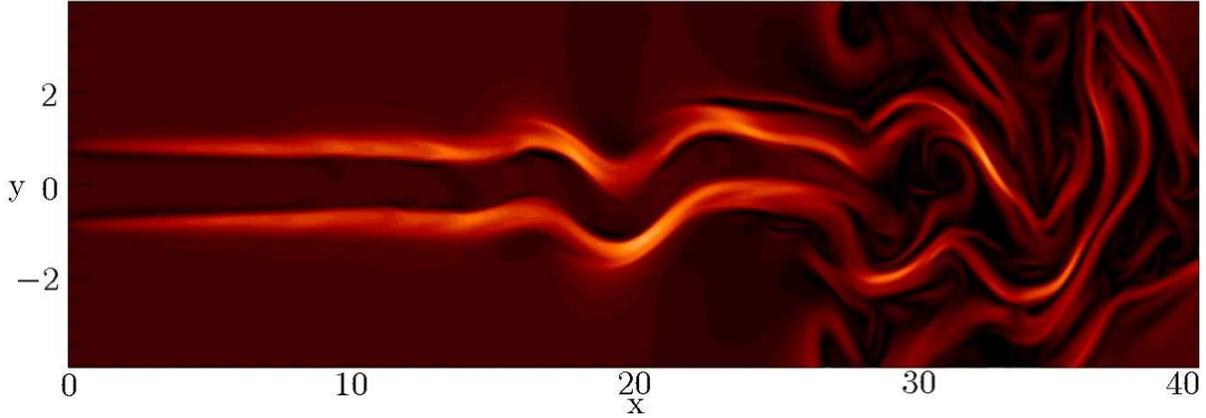}
      \caption{Magnetic intensity distribution in run 2 at $t = 80$. A linear scale is used with 
magnetic intensity ranging from 0. (dark) to 1. (bright). The unit of length is the jet diameter.}
      \label{t4mag}
   \end{figure*}   

However, our spatial simulations indicate that a different scenario occurs at a later time. This is 
obvious in Fig. \ref{jetl80} (at $t = 320$), showing that the jet flow has been revived after 
the large-scale disruption. This figure was obtained from run 6 when the dimension of the domain 
was twice as long compared to our standard run 2, in order to investigate the large-scale behavior of the jet.
In Sect. \ref{overview}, we proposed the presence of a trail structure propagating in the upstream direction, as shown for example in the density map of Fig 3.

\begin{figure}[t]
   \centering
   \includegraphics[angle=-90,width=9cm]{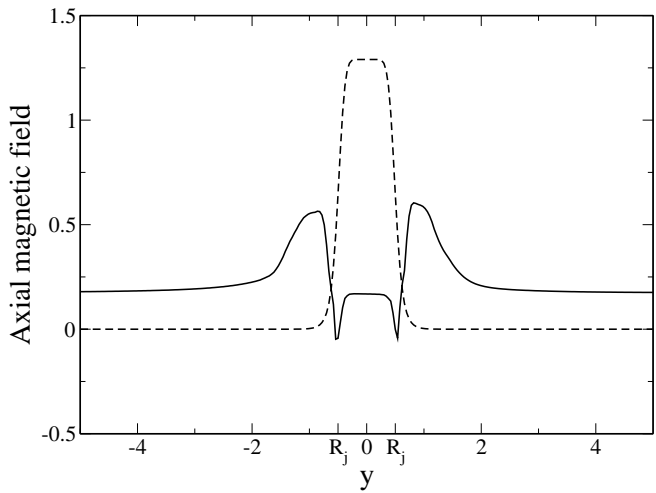}
      \caption{One dimensional cut at $x = 3$ of the axial magnetic field in Fig. \ref{t4mag} (plain line). The velocity profile is overplotted for comparison (dashed line).}
         \label{fixed_bc}
 \end{figure}
 
This behavior correlates to a local strengthening of the magnetic field. This is clearly visible in Fig. \ref{t4mag}, which shows a map of the magnitude of the magnetic field for run 2 at time $t=80$ where the trail perturbation has just reached the jet inlet. In addition, Fig. \ref{fixed_bc}  shows the corresponding transverse profile of the longitudinal magnetic field. The cut is made close to the inlet boundary (at $x = 3$), but a similar shape is also obtained along nearly the whole jet at this time (i.e. for $x \lesssim 20$). As one can see in the latter figure, slightly outside the jet's radius the magnetic field is amplified by a factor up to $\sim 3.5$ (i.e. $\sim M_A/2$) compared to its initial value. This is not true in the jet core, where the magnetic field has only slightly decreased. This new magnetic structure is observed to be maintained in time, and thus it changes the stability of the jet flow at later times.

The question remains, what is the physical origin of this amplification? It has previously been shown that during the linear and early non-linear evolution, the KH vortices (here of the instability train) are able to expel the magnetic field from the vortex center, stretching it and amplifying it around the vortex perimeter (e.g. Jones et al. 1997). A non-linear saturation then occurs when the magnetic field becomes locally dominant, i.e. when the field line tension is able to overcome the centrifugal force associated with the vortical motion. In our spatial simulations, we observe that the corresponding magnetic perturbation is able to travel backward up to the jet inlet, giving rise to the trail seen before. This is possible because the magnetic perturbation is traveling at the local Alfv\'en speed in the jet's frame which has a local speed close to zero (see Fig. 11 at $y\approx 1$).

\subsubsection{Influence of $M_A$}

\begin{figure}
   \centering
   \includegraphics[angle=-90,width=9cm]{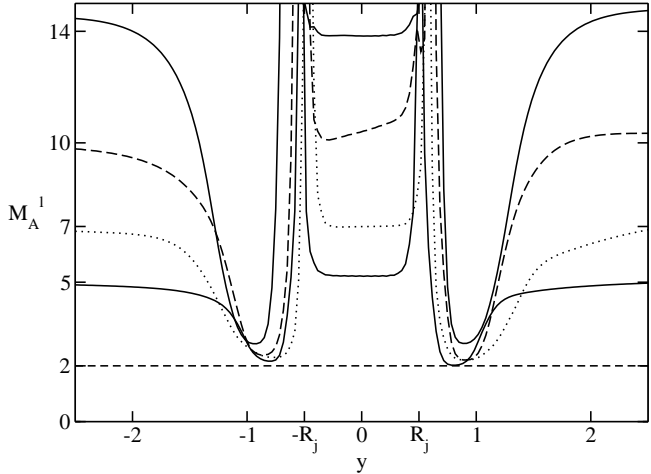}
      \caption{One dimensional cut at $x=3$ of the local Alfv\'enic Mach number  $M^l_A$ (see text) for 
different initial Alfv\'enic Mach numbers $M_A = 5, 7, 10, 14$ (from the bottom to the top) corresponding to the transonic runs 1, 2, 3 and 4 (see Tab. 1).}
         \label{profils_ma}
 \end{figure}

We investigated the influence of the choice of the initial magnitude of the magnetic field 
(measured using the Alfv\'en Mach number $M_A$) on the final configuration.
The results obtained for the final magnetic profiles are plotted in Fig. \ref{profils_ma} using the local 
Alfv\'enic Mach number defined as:

\begin{equation}
M_A^l = V_0/V_A(y) 
\label{localma}
\end{equation}

\noindent where $V_0$ is the velocity of the jet and $V_A(y) = B_x(y)/\sqrt{\rho(y)}$ is the local Alfvenic speed.
This number is similar to the initial $M_A$, but uses the local value of the magnetic field intensity and plasma density. 
Figure \ref{profils_ma} shows that a value close to $2$ is locally reached, irrespective of the initial $M_A$ value. This means that the field is amplified up to a local maximum value corresponding to $M^l_A = 2$. Consequently, the amplification mechanism is rather robust as it works in a similar way for a rather large range of initial magnetic field amplitudes for transonic jets.

\subsubsection{Stability analysis of the final configuration}

A single shear flow layer embedded in a uniform magnetic field with an Alfv\'en Mach number smaller than $2$ is known to be fully linearly stable. However, our final 2D slab jet configuration is different (i.e. a double shear flow layer with a non-uniform magnetic field) and thus requires a stability analysis.

\begin{figure}[t] 
   \centering
   \includegraphics[width=9.5cm]{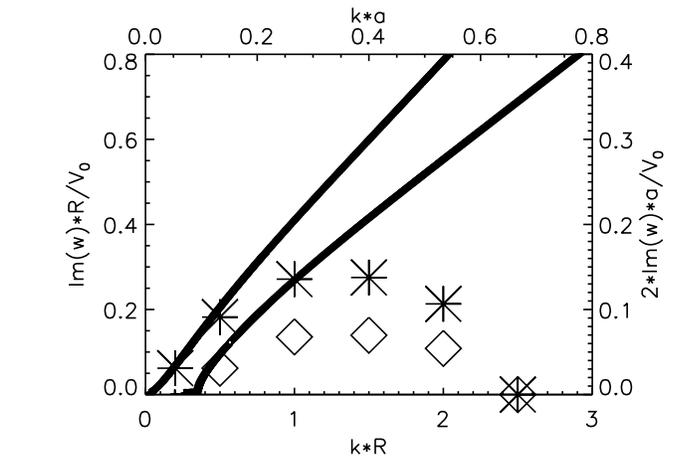}
   \caption{Temporal growth rate curves obtained by solving Hardee et al. (1992) dispersion relation with the vortex sheet
   approximation, for the initial equilibrium (upper curve)
   and  for the final equilibrium (lower curve).  The growth rates obtained using VAC code for 5 wavenumbers are also plotted
   (stars and diamonds are used for initial and final equilibria respectively).}
   \label{gr_curves}
\end{figure}

We investigated the solutions of the dispersion relation derived by Hardee et al. for a magnetized slab jet (i.e. given by Eq. (1) in Hardee et al. 1992).
An idealized equilibrium state is assumed, as the latter dispersion relation requires an equilibrium configuration with a vortex sheet (i.e. a discontinuity between the jet flow and the external medium). First, as an equilibrium state is not strictly reached at the end of our spatial simulation (see below), we calculated a neighboring equilibrium having similar physical parameters.
Second, we consider uniform magnetic fields inside the jet core (subscript ``in") with 
$M_\mathrm{A,in}=7$ and in the external medium (subscript ``ex")  with $M_\mathrm{A,ex}=2$. The density ratio $\eta = \rho_\mathrm{in}/\rho_\mathrm{ex}$ is taken to be unity and the sonic Mach numbers are consequently $M_\mathrm{s,in} = 1$, and $M_\mathrm{s,ex} = 1.11$.

The results are plotted in Fig. 13 for the temporal growth rate of the antisymmetric mode (which is the most unstable one). For comparison, the same stability curve for our initial uniformly magnetized configuration was also computed (in this case the parameters are $\eta = \rho_\mathrm{in}/\rho_\mathrm{ex} = 1$, $M_\mathrm{s,in} = M_\mathrm{s,ex} =1$, and $M_\mathrm{A,in} = M_\mathrm{A,ex} = 7$). One can see the enhanced stability effect for the final configuration (compared to the initial one) due to the enhanced magnetization of the external medium. However, a full stability is not achieved. As the latter stability study does not take into account the effect of the smooth transition velocity profile at the jet interface (over a length scale $a=0.125$), we additionally plotted the temporal growth rate of five wavelengths, measured from additional temporal runs where a single wavelength is driven unstable.
The values obtained agree with the results deduced from the vortex sheet curves for wavelength $\lambda \gg a$ (or equivalent for small enough wavenumbers).
However, small wavelengths (or equivalently $ka \lesssim 0.7$) become stable as expected from the effect of velocity gradients (Ferrari et al. 1982).
The linear stability enhancement between the two equilibria is of the order of 50 percent, as measured when taking the wavelength of the fastest linearly growing mode.

\begin{figure*}[t]
   \centering
   \includegraphics[width=18cm]{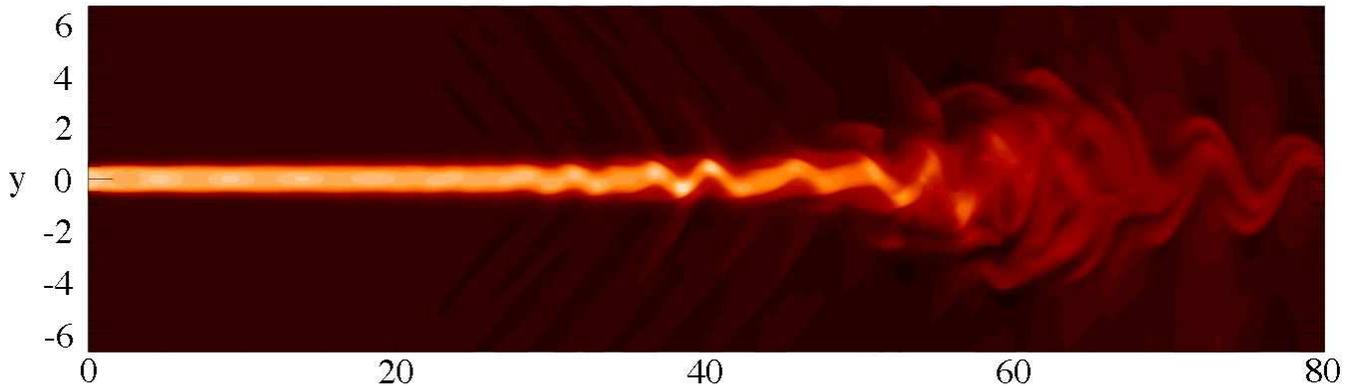}
      \caption{Axial velocity distribution in the supersonic case (run 8 : $M_s = 3$, $M_A = 7$) at $t = 300$. The unit of length is the jet diameter.}
         \label{jetl80ms3}
 \end{figure*}
 
Thus, according to our analysis, the linear stability of the final jet configuration is enhanced by a factor of approximately two due to the presence of a relatively strongly magnetized external medium (with a local Alfv\'en Mach number of order 2). However, this is only a partial stabilization, as unstable modes with non-negligible growth rates remain present. This is in agreement with the relatively short wavelengths seen at $x \approx 20$ in Fig. \ref{jetl80}. Note that these short wavelength modes do not evolve into KH vortices, i.e. the non-linear evolution of the instability is different in the new configuration.
 
As is the case for the residual large-scale oscillation of the whole jet with wavelength $\lambda \sim 30$ (see Fig. 9 at $x\approx 30$), it is unlikely that this is due to the development of KH modes as our stability analysis predicts full stability for the corresponding wavenumber $ka \sim 0.02$. In fact, we have checked that, at the end of the simulation, an exact equilibrium is not yet fully reached and the system therefore oscillates when trying to relax towards it.

\subsection{Supersonic case}
\label{supersonic}
 
We also performed a jet simulation having a supersonic flow with a Mach number $M_s = 3$ 
(run 8 in Table 1) and show an axial velocity snapshot in Fig. \ref{jetl80ms3}. Two essential 
differences are evident when compared to the transonic regime. First, the jet perturbation is characterized by sinusoidal oscillations of the whole flow and not by vortex-like structures (observed in transonic jets).
This is due to the fact that the modes change character when $M_f  \gtrsim 2$ (in our case $M_f = 2.75$, Paper I). 
We deduce from the periodicity of the observed sinusoidal features that the wavelength of the fastest growing
mode is $\lambda \approx 4$ in our units (giving thus $ka \approx 0.2$), in agreement with previous linear stability results (see lower curve in Fig. 2 of Paper I).
In the latter stability study, the dominant mode is shown to be a KH surface mode.
As concerns the growth rate, we deduce a value smaller than that obtained from the transonic run. This is not surprising as the difference is mainly due to the fact that a perturbation convected by a supersonic flow is able to propagate further than by a transonic flow during the same given time. KH modes in a transonic flow (giving rise to two rows of vortices at the interface) translates into ``surface" KH modes (in a supersonic flow) having  longer wavelengths and lower spatial growth rates. 
In addition to the surface mode, internal modes become unstable in the supersonic regime (again when $M_f \gtrsim  2$).
These ``body" modes are typical of two-shear and cylindrical layer configurations, as they become unstable by resonant reflection at the jet boundary (Ferrari 1998; Birkinshaw 1991). The latter modes are not expected to be dangerous for the jet survival.
In our case, the dominant body modes are expected to have a wavelength $\lambda \lesssim 1$ (in our units) and a rather low growth rate in comparison to the surface mode (see lower curve in Fig. 2 of Paper I), thus explaining why they are not visible in Fig. \ref{jetl80ms3}.

The long-term evolution of our run shows that the previously observed stabilization effect of the 
jet is now absent. This is not surprising, as the mechanism of magnetic field enhancement driven by 
the vortical deformation of the interface is not present for the supersonic flow. We also observe that 
the jet is disrupted in a similar way as obtained in temporal runs of Paper I, where shock dominated 
transients characterize the disruption.
Moreover, a propagation distance of order $100 R_j$ is observed in  this supersonic case. 

\section{Conclusion}
\label{conclusion}
We can summarize our findings as follows. We have numerically investigated the spatial development of the KH instability in jets embedded in uniform magnetic fields. We focus mainly on the transonic regime for weakly magnetized plasmas, for which the plasma-beta is much greater than unity.

First, we confirm that a temporal approach provides a good approximation for modeling the
initial stages of development of the KH modes. The latter assertion comes from a detailed comparison between our spatial runs and previous/additional temporal simulations. This is true for early times corresponding to the linear and early non-linear phases, ending up with a large-scale disruption of the flow.

However, at later times the evolution differs drastically in our spatial vs temporal runs. Indeed, a temporal run is known to definitively terminate in a broadened and heated residual flow. As a consequence, a temporal transonic jet flow is not able to survive on an equivalent distance exceeding approximately 40 times the jet radius. On the other hand, our transonic spatial run clearly shows that the initial jet flow is able to revive and maintain its coherence at a later time after the large-scale disruption. A corresponding partial stabilization mechanism is identified. Indeed, the primary KH modes induce the formation of a sheath region of enhanced magnetic field situated around the jet core. This region is an envelope that primarily coincides with the region where KH vortices saturate, and progressively extends itself in the upstream direction towards the jet inlet. In this way, the stability of the whole flow structure appears to become significantly enhanced, as local Alfv\'en Mach values between $2$ and $4$ are reached in this region. This stabilization mechanism is not at work for supersonic flows (rolling-up vortices are absent) which consequently do not survive KH instabilities. In the supersonic regime, the temporal and spatial approaches lead to the same conclusion: a rapid shock dominated disruption terminates the jet flow.

Pioneering studies on spatial simulations of the development of KH instabilities in 2D unstable jets were performed by Hardee et al. (Hardee et al. 1992, 1995). However, these studies have focused on a different magnetic regime, as rather strongly magnetized configurations with a plasma-beta of order unity were considered.
In a real jet (YSO or AGN for example), the plasma parameters are not known even if one expects an average plasma-beta of order unity. For example, it cannot be excluded that
a high value of the plasma-beta of order 100 is reached in a central region of the jet core (see Fig. 6 in Lery et al. 2000).
Thus, before reaching the asymptotic region where the flow is supersonic, the plasma of a real jet is accelerated in an intermediate transonic region. The partial stabilization mechanism presented in this study could thus help to explain the jet survival in the transonic regime. This is possible if the transonic phase lasts long enough for the stabilization effect to operate. In the opposite case, the flow is anyway stable as the KH modes do not have enough time to grow. For example in the accretion/ejection model of Casse \& Keppens (2002, 2004), velocity curves show that the jet is accelerated into the supersonic regime on a very short-length scale, of the order $50$ times the inner radius of the accretion disc. As the jet in their simulation has a radius of approximatively $20$ inner radius, it shows that according to this model the transonic phase has a very short duration.
Unfortunately, our stabilization mechanism does not work in the supersonic regime, for which the problem of the jet survival remains unsolved in spite of many recent attempts (Paper I; Hardee \& Rosen 2002; review by Hardee 2004).

Note also that in the present study, we use idealized equilibrium profiles for the jet configuration to facilitate the interpretation of the results. It would be interesting to test the robustness of the  stabilizing mechanism in the presence of more realistic profiles. For example, the velocity profile expected in astrophysical jets is probably much more complex, consisting of different components (e.g. Frank et al. 2000). This latter interesting question is left to a future investigation.

Finally, it will be of interest to study whether the results obtained in this study apply also in the 3D case. We do not expect strong differences, mainly because there is a certain correspondence in the mode nomenclature between 2D and 3D configurations. The only known difference is the enhanced turbulent aspect in 3D due to the presence of typical 3D modes and non-linear couplings (Paper I).

\begin{acknowledgements}
The numerical simulations in this paper were performed by the Versatile Advection Code maintained by G. T\' oth and R. Keppens (see http://www.phys.uu.nl/~toth/). Numerical calculations were performed on the SGI Origin 3800 at CINES, Montpellier (France). We thank the anonymous referee for interesting suggestions that helped improve the paper.

\end{acknowledgements}


\begin{thebibliography}{}

  \bibitem[1992]{appl1} 
  Appl, S., Camenzind, M. 1992
  A\&A, 256, 354
  
  \bibitem[1996]{appl2} 
  Appl, S. 1996
  A\&A, 314, 995
  
  \bibitem[2003]{baty2003} 
  Baty, H., Keppens, R., Comte, P. 2003,
  Phys. Plasmas, 10, 4661
  
  \bibitem[2005]{baty2005} 
  Baty, H. 2005
  A\&A, 430, 9
  
  \bibitem[2006]{baty2006} 
   Baty, H., Keppens, R. 2006,
   A\&A, 447, 9 (Paper I)
      
  \bibitem[1995]{bodo1}
  Bodo, G., Massaglia, S., Rossi, P., Rosner, R., Malagoli, A., Ferrari, A., 1995
  A\&A, 303, 281
  
  \bibitem[1998]{bodo2}
  Bodo, G., Rossi, P., Massaglia, S., Ferrari, A., Malagoli, A.,  Rosner, R.,  1998
  A\&A, 333, 1117
  
  \bibitem[1991]{birkinshaw} 
      Birkinshaw, M. 1991,
      in Beams and Jets in Astrophysics, ed. P.A. Hughes
      (Cambridge Univ. Press), 279

   \bibitem[2002]{casse1}
   Casse, F., Keppens, R. 2002,
   ApJ, 581, 988
 
   \bibitem[2004]{casse2}
   Casse, F., Keppens, R. 2004,
   ApJ, 601, 90
   
   \bibitem[1984]{colella}
   Colella, P., Woodward, P.R. 1984,
   J. Comput. Phys., 54, 174
    
   \bibitem[1981]{drazin}
   Drazin, P.G., Reid, W.H. 1981,
   Hydrodynamic Stability, Cambridge Univ. Press

   \bibitem[2000]{frank}
   Frank, A., Lery, T., Gardiner, T.A., Jones, T.W., Ryu, D. 2000
   ApJ, 540, 342
   
    \bibitem[1982]{ferrari}
   Ferrari, A., Massaglia, S., Trussoni, E. 1982
   MNRAS, 198, 1065

   \bibitem[1998]{ferrari2}
   Ferrari, A., 1998
   Annual Review of Astronomy and Astrophysics, Volume 36, pp. 539-598
   
   \bibitem[2004]{gabuzda}
   Gabuzda, D.C., Murray, E., Cronin P. 2004,
   MNRAS, 351, 89
      
  \bibitem[1992]{hardee1} 
      Hardee, P.E., Cooper, M. A., Norman, M. L., Stone, J. M. 1992,
      ApJ, 399, 478

  \bibitem[1995]{hardee1b} 
      Hardee, P.E., Clarke A. C. 1995,
      ApJ, 449, 119

  \bibitem[1995]{hardee1c} 
      Hardee, P.E., Clarke A. C., Howell D.A. 1995,
      ApJ, 441, 644
	
  \bibitem[2002]{hardee2} 
      Hardee, P.E., Rosen, A. 2002
      ApJ, 576, 204
  
  \bibitem[2004]{hardee3}
     Hardee, P.E. 2004
     Ap\&SS, 293, 117
     
  \bibitem[1983]{harten} 
  Harten, A. 1983,
  J. Comp. Phys., 49, 357
              
  \bibitem[1997]{jones} 
     Jones, T. W.; Gaalaas, Joseph B.; Ryu, Dongsu; Frank, Adam
     1997, ApJ, 482, 340
    
  \bibitem[2004]{keppens1999} 
    Keppens, R., T\'oth G., Westermann H. J., Goedbloed, J.P. 1999
     J.Plasma Physics, 61, 1

  \bibitem[2000]{lery2000} 
     Lery, T., Frank, A.
     ApJ, 533, 2000
     
   \bibitem[1998]{micono} 
      Micono, M., Massaglia, S., Bodo, G., Rossi, P., Ferrari, A 1998,
      A\&A, 333, 989

   \bibitem[1997]{min} 
      Min, K.W. 1997
      ApJ, 482, 733

   \bibitem[1983]{miura} 
	Miura, A., Pritchett, P.L. 1983
	JGR, 87, 7431
	
   \bibitem[1985]{payne} 
      Payne, D. and Cohn, H. 1985,
      ApJ, 291, 655

  \bibitem[1997]{ray}
   Ray, T.P., Musclow, T.W.B., Axon, D.J., et al. 1997,
   Nature, 385, 415
   
   \bibitem[1996]{toth} T\'oth, G. 1996, Astrophysical Letters \& Communications, 34, 245

   \bibitem[2005]{push}
   Pushkarev, A.B., Gabuzda, D.C. Vetukhnovskaya, Y.N., Yamikov, V.E. 2005,
   MNRAS, 356, 859
   
   \bibitem[yamamoto]{yamamoto}
   Yamamoto, T., Yamashita., M.A. 1988,
   Phys. Fluids, 31, 2152

   \bibitem[1992]{zhao1992}
   Zhao, J.H., Burns J.O., Normal, M.L., Sulkanen, M.E. 1992
   ApJ, 387, 83
   
\end{thebibliography}
\end{document}